\documentstyle[11pt,aaspp4]{article}

\def \m{\ifmmode M_\odot\else M$_\odot$\fi}
\def \r{\ifmmode R_\odot\else R$_\odot$\fi}

\def\gmcm3{g~cm$^{-3}$}
\def\gm-s{g~s$^{-1}$}
\def\cm3s{cm$^3$~s$^{-1}$}
\def\kms{km~s$^{-1}$}
\def\erg-s{erg~s$^{-1}$}

\def\beq{\begin{equation}}
\def\eeq{\end{equation}}
\def\ref{\reference}
\def\gr{$\gamma$-ray}

\def\grb{$\gamma$-ray burst}

\def\0{\parindent=0cm}
\def\5{\parindent=.5cm}
\def\7{\parindent=.7cm}

\begin{document}

\title{\bf Jet-Induced Explosions of Core Collapse Supernovae}
\author{A.M. Khokhlov$^1$, P.A. H\"oflich$^2$,  E.S. Oran$^1$,
J.C. Wheeler$^2$ \& L. Wang$^2$}
\affil{$^1$Laboratory for Computational Physics and Fluid Dynamics, \\
Naval Research Laboratory,
Washington, DC 20375\\           
ajk@lcp.nrl.navy.mil, oran@lcp.nrl.navy.mil}
\affil{$^2$Astronomy Department, University of Texas, Austin, Texas 78712\\
pah@alla.as.utexas.edu, wheel@astro.as.utexas.edu, lifan@tao.as.utexas.edu}

\begin{abstract}             

We numerically studied the explosion of a supernova caused by
supersonic jets present in its center. The jets are assumed to be
 generated  by a magneto-rotational mechanism when a
stellar core collapses into a neutron star. We  simulated  the process of
 the
jet propagation through the star, jet breakthrough, and the ejection of
the supernova envelope by the  lateral shocks generated during jet
propagation. The end result of the interaction is a highly nonspherical
supernova explosion with two high-velocity jets of material moving in
polar directions, and a slower moving, oblate, highly distorted 
ejecta containing most of the supernova material. 

The jet-induced explosion is entirely due to the action of the jets on the
surrounding star and does not depend on neutrino transport or
re-acceleration of a stalled shock.
The jet mechanism can explain the observed high polarization of Type 
Ib,c and Type II supernovae, pulsar kicks, very high velocity material
observed in supernova remnants, 
 indications that radioactive material was carried  to the hydrogen-rich
layers in SN1987A, and some others observations that are very difficult
 or impossible to explain by the neutrino energy deposition mechanism.
The breakout of the jet from a compact,
hydrogen-deficient core may account for the
gamma-ray bursts and radio outburst associated with SN~1998bw/GRB980425.

\end{abstract}             

\keywords{supernovae: general; individual (SN~1998bw) $-$ gamma rays: bursts 
$-$ pulsars:general $-$ ISM: jets and outflows}

\section{Introduction}

Recent observations of core collapse supernovae provide increasing
evidence that the core collapse process is intrinsically asymmetric:

1) The spectra of these supernovae are significantly polarized indicating
 asymmetric envelopes (M\'endez et al. 1988; H\"oflich 1991, 1995; Jeffrey
1991; Trammel et al. 1993, Tran et al. 1997).
The degree of polarization  tend to vary inversely with the mass of the hydrogen
envelope, being maximum for Type Ib/c events with no hydrogen (Wang et
al. 1996; Wang, Wheeler \& H\"oflich 1999; Wheeler, H\"oflich \& Wang
1999).

2) After the explosion, neutron stars are observed with high
velocities, up to 1000 \kms (Strom et al. 1995).

3) Observations of SN~1987A showed that radioactive material was brought
to the hydrogen rich layers of the ejecta very quickly during the explosion
(Lucy 1988; Sunyaev et al. 1987, Tueller et al. 1991).

4) The remnant of the Cas A supernova shows rapidly moving oxygen-rich
matter outside the nominal boundary of the remnant  (Fesen \&
Gunderson, 1996) and  evidence for two oppositely directed jets of
high-velocity material (Fesen 1999; Reed, Hester, \& Winkler 1999). 

5) High velocity ``bullets" of matter have been observed in the Vela
supernova remnant (Taylor et al. 1993)

Understanding the mechanism of producing supernovae explosions by  core
collapse is a physics problem that has challenged researchers for
decades (Hoyle \& Fowler 1960; Colgate \& White 1966).   The current
most sophisticated calculations based on the neutrino energy
deposition mechanism are multidimensional and involve the convection
of the newly formed neutron star. These, however,  but have failed to
produce robust explosions (Herant et al. 1994; 
Burrows, Hayes \& Fryxell 1995; Janka \&
M\"uller 1996; Mezzacappa et al 1997; Lichtenstadt,
Khokhlov \& Wheeler 1999). Even when successful, these models do not  
explain why  SN 1998bw produced the strongest 
radio source ever associated with
a supernova, probably requiring a relativistic blast wave (Kulkarni
et al. 1998), or account for a probable link between SN 1998bw and 
the \grb\ GRB980425
observed in the same general location in the same general time frame
(Galama et al. 1998). 

The discovery of pulsars led to early
considerations of the role of rotating magnetized neutron stars in the
explosion mechanism (LeBlanc and Wilson 1970; Ostriker \& Gunn 1971; Bisnovatyi-Kogan 1971). 
LeBlanc and Wilson studied the magneto-rotational core collapse
of a $7 M_{\odot}$ star. They numerically solved the two-dimensional
MHD equations coupled to the equation for neutrino transport. Their
simulations showed the formation of two oppositely directed,
high-density, supersonic jets of material emanating from the collapsed
core. They estimated that at the surface located $\sim 4 \times
10^8$~cm from the center, the jet carried away $\sim 10^{32}$~g  with
$\sim 1-2 \times 10^{51}$~ergs in $\sim 1$~s.   The magnetic field
generated in this calculation was $\sim 10^{15}$ Gauss. Evidence now
exists for strongly magnetized neutron stars, ``magnetars" (Duncan \&
Thompson 1992; Kouveliotou et al. 1998).
  
The LeBlanc-Wilson mechanism is extremely asymmetric and contains jets.
Their calculations only followed the jet to a  distance of $\sim 10^8$
cm, whereas a stellar core has a radius of $10^{10}$~cm or more. The
issues that arise are: how can this  asymmetry propagate to much larger
distances inside the star? Can these jets induce asymmetry at distances
comparable to the stellar radius, or even push through the entire star
and exit? 

In this paper, we model the explosion of a core collapse supernova
assuming that the LeBlanc-Wilson mechanism has operated in the center. 
We take a $15 M_{\odot}$ main-sequence star evolved to
the point of the explosion (Straniero, Chieffi \& Limongi 1999) and assume that the 
star has lost all of its hydrogen envelope before the explosion.  The
resulting $4.1$\m\ model of a helium star 
 corresponds to the  explosion of a Type Ib or Ic
supernova.  The simulations show that the jets cause a very asymmetric
explosion of the star.  Most of the observations of
asymmetries listed above can be explained by this process.

\section{Numerical Simulations}

Figure 1 presents a schematic of the setup of the computation.   The
computational domain is a cube of size $L = 1.5 \times 10^{11} $~cm
with a spherical helium star of radius $R_{\rm star} =  1.88 \times
10^{10}$~cm and mass $M_{\rm star}\simeq~4.1$\m\  placed in the center. 
The distribution of physical parameters inside the star is shown in
Figure 2.  The innermost part with mass $M_{\rm core} \simeq 1.6 
M_{\odot}$ and radius  $R_{\rm core} = 3.82 \times 10^8$~cm, consisting
of Fe and Si, is assumed to have collapsed on a timescale much faster
than the outer, lower-density material. It is removed and replaced by a
point gravitational source with mass $ M_{\rm core}$ representing the
newly formed neutron star.  The remaining mass, from $\simeq 1.6$ to
$\simeq 4.1 M_{\odot}$, consists of an O-Ne-Mg inner layer surrounded by
the C-O and He-envelopes.  This structure is mapped onto the
computational  domain from $R_{\rm core}$ to $R_{\rm star}$. 

At $R_{\rm core}$  and the outer boundary of the computational domain,
we impose  an outflow boundary condition assuming zero pressure,
velocity, and density gradients. At two polar locations where the jets
are initiated at $R_{core}$, we impose an inflow with velocity $v_j$, density
$\rho_j$ and pressure $P_j$.   The jet parameters are chosen to
represent the results of  LeBlanc \& Wilson (1970). At  $R_{\rm core}$,
the jet density and pressure are the same as those of the background
material, $\rho_j = 6.5 \times 10^5$~\gmcm3 and $P_j = 1.0 \times
10^{23}$~ergs~cm$^{-3}$, respectively. The radii of the  cylindrical jets
entering the computational domain  are approximately $r_j = 1.2 \times
10^8$~cm.
 
For the first 0.5~s, the jet velocity at $R_{\rm core}$ is kept
constant at $v_j = 3.22 \times 10^9~cm~s^{-1}$. This results in a mass flux
rate of $\sim 9.5\times10^{31}$ \gm-s with an energy deposition rate $d
E / dt = 5\times 10^{50} $~ergs/s for each jet.  After 0.5~s, the
velocity of the jets at $R_{\rm core}$ was gradually decreased to zero
at approximately 2~s. The total energy deposited by the jets is $E_j
\simeq 9\times 10^{50}$~ergs and the total mass ejected is $M_j \simeq
2\times 10^{32}$~grams or $\simeq 0.1$\m. These parameters are
consistent within, but somewhat less than, those of the LeBlanc-Wilson model.
The amount of material ejected is less than that which falls through
the inner boundary during the jet operation, $\simeq 4\times10^{32}$~g.
This amounts to an implicit assumption that $\sim 1/2$ of the matter
accreted is channeled back out into the
jets.  More accurate jet parameters can only be determined by
self-consistently modeling 
the formation of the jets in the vicinity of a neutron star.

The stellar material was described by the time-dependent, compressible,
Euler equations for inviscid flow with an ideal gas equation of state 
$P=E(\gamma-1)$ with constant $\gamma=5/3$. The Euler
equations were integrated using an explicit, second-order accurate, Godunov type,
adaptive-mesh-refinement, massively parallel, Fully-Threaded Tree (FTT)
program, ALLA  (Khokhlov 1998, Khokhlov \& Chtchelkanova 1999). Euler
fluxes were evaluated by solving a Riemann problem at cell interfaces.
FTT discretization of the computational domain allowed the mesh to be
refined or coarsened at the level of individual cells. Physical scales
involved in the simulation range from the size of the computational domain
($1.5 \times 10^{11} $~cm) to the jet diameter ($ \sim 10^8$~cm) and
span at least three orders of magnitude.  We used a cartesian,
nonuniformly refined FTT mesh with   fine cells $\Delta_{\rm min}\simeq
3.7\times 10^7$~cm near $R_{\rm core}$ to resolve the jets, and with
cell size increasing towards the outer boundary of the computational
domain where the cell size was $\Delta_{\rm max} = 2.3\times 10^9$~cm. 
This mesh was fixed from initial time $0$ to $ 6$~s of physical time.
After that, the inner parts were coarsened near the center by a factor
of four, and the central hole was eliminated. At this moment, the jets
have exited the star and the details of the flow near $R_{\rm core}$ do
not affect the essential features of the explosion. In this first,
demonstration calculation, we did not use the time-adaptive mesh
refinement capability of ALLA.  It will be used to follow shocks and
mixing processes with higher resolution in future simulations. We 
computed the entire configuration including both jets and assuming no
symmetries. The total number of computational cells used in the
simulation was $\sim 2\times 10^6$, whereas a uniform resolution
$\Delta_{\rm min}$ would have required $\sim 7\times 10^{10}$ cells.

\section{Results and Discussion}

Figure 3 shows the propagation of the jet inside the star. As the jets
move outwards, they remain collimated and do not  develop much internal
structure. A bow shock forms at the head of the jet and spreads in all
directions, roughly cylindrically around each jet.   The sound crossing
time $\tau(r) = r/ a_s(r)$  is shown as a function of stellar radius
$r$ in Figure 2, where $a_s(r)$ is the sound speed at a given radius
for the initial stellar model.  It  might be expected that if energy
were released at the center of a star on a timescale much shorter that
$\tau(r)$, the effect of energy deposition at $r$ would resemble that
of a strong point explosion. In particular, the jet characteristic time
$\tau_j\sim 1$~s is  much shorter than the sound crossing time of the
star, $\tau(R_{\rm star}) \sim 10^3$~s (Figure 2).  Nonetheless,
these  jets stay collimated enough to reach the surface as strong jets. 

 It is known that supersonic
 jets stay collimated for a long distance. For example, Norman et al.
(1983) simulated supersonic jets  with densities  $\rho_j$ both less than
and greater than a uniform background, $\rho_b$. Jets with
$\rho_j/\rho_b \ge 1$ developed a bow shock and little internal
structure. Our jets resemble those with $\rho_j/\rho_b\ge 1$. The
stellar matter is shocked by the bow shock, and then flows  out and
acts as a high-pressure confining medium by  forming a cocoon around
the jet. The sound crossing time of the dense O-Ne-Mg mantle,
$\tau(R\sim 10^9~{\rm cm}) \simeq 10$~s, is only ten times longer than
$\tau_j$, and the jets are capable of penetrating this dense inner part
of the star in $\sim 2$~s.  By the time the jets penetrate into
the less dense C-O and He layers, the inflow of material into the jets has
been turned off. By this time, however, the jets have become long
bullets of high-density material moving through the background
low-density material almost ballistically. The higher pressures in
these jets cause them to spread laterally. This spreading is limited by
a secondary shock that forms around each jet  between the jet  and the
material already shocked by the bow shock.   The radius of the jets,
$\sim 3\times10^9$ cm as they emerge from the star, is larger than the
initial radius, $\sim 10^8$ cm,  but it is still significantly less
than the radius of the star. 

After about 5.9 s, the bow shock reaches the edge of the star and
breaks through. Figure 4 shows the subsequent evolution of  the star
after the breakthrough.  By $\simeq 20$~s, most of the material in the
jets has left the star propagates into the interstellar medium 
ballistically. We estimate the total mass in these two  jets as $M_j
\approx 0.05 M_\odot$ and the total kinetic energy $E_j \approx 2.5 \times
10^{50}$~ergs. The average velocity of the jet is about 25,000~$km~s^{-1}$.

The laterally expanding bow shocks generated by the jets (Figure 3) move
towards the equator where they collide with each other. The collision
of the shocks first produces a regular reflection that then becomes a
Mach reflection. The Mach stem moves outwards along the equatorial
plane. The result is that the material in the equatorial plane is
compressed and accelerated more than  material in other directions
(excluding the jet material). At $t\simeq 29$~s, the Mach stem reaches
the outer edge of the star, and the star begins to settle into the free
expansion regime. The computation was terminated at $\simeq 35$~s,
before free expansion was attained. The stellar ejecta at this time is
highly asymmetric. The density contour of $50~{\rm g~cm^{-3}}$, which is
the average density of the ejecta at this time, forms an oblate
configuration with the equator-to-polar velocity ratio 
$\simeq 2/1$. Complex shock and rarefaction
interactions inside the expanding envelope will continue to change the
distribution of the parameters inside the ejecta. Nonetheless, we
expect that the resulting configuration will resemble an oblate
ellipsoid with a very high degree of asymmetry, axis ratios $\ge 2$.

\section{Conclusions}

We have numerically studied the explosion of a supernova caused by
supersonic jets generated in the center of the supernova as  a result of
the core collapse into a neutron star. We  simulated  the process of the
jet propagation through the star, jet breakthrough, and the ejection of
the supernova envelope by the  lateral shocks generated during jet
propagation. The end result of the interaction is a highly nonspherical
supernova explosion with two high-velocity jets of material moving in
polar directions ahead of an oblate, highly distorted ejecta containing
most of the supernova material.
Below we argue that such a model explains many of the observations that
are difficult or impossible to explain by the neutrino deposition
explosion mechanisms.
 
We have assumed that the jets were generated by a magneto-rotational
mechanism during core collapse and neutron star formation (LeBlanc \&
Wilson 1970).  That collimated jets could be a common phenomenon in 
core collapse supernovae and be associated with \gr\ bursts was raised
recently by Wang \& Wheeler (1998). A different  mechanism of jet
generation involving neutrino radiation during collapse of a very
massive star into a black hole has been recently discussed by MacFayden \&
Woosley (1999) in the context of a ``failed'' supernova,
 also to explain \gr\ bursts. 
Low density relativistic jets may also be
produced by the intense radiation of the newly born pulsar, as discussed
by Blackman \& Yi (1998) and Yi et al. (1999). 
 We found in our preliminary simulations (not presented here)
that lower density and higher velocity jets than the one considered in
this paper may produce similar hydrodynamical effects. 

The asymmetric explosion generated in this calculation provides
ejection velocities that are comparable to those observed in
supernovae.  For this particular calculation,  an energy of 
$2.5\times 10^{50}$ ergs is invested in the jet and the star
of $\simeq 2.5$\m\ is ejected with kinetic 
energy of $6.5\times 10^{50}$ ergs and average velocity 
$3,000-4,000~km~s^{-1}$. 
Increasing the jet opening angle, jet duration,
 or jet velocity would result in a more powerful explosion.
The density and velocity profiles of 
the main ejecta (excluding jets) are oblate with equator to polar ratios
greater than 2/1.
%
%
This structure will produce significant polarization, of order
1\% or more as observed in bare-core supernovae
(H\"oflich, Wheeler \& Wang 1999).

The two polar jets move outward from the star with a
 speed $\sim 25,000~km~s^{-1}$,
much greater than the ejecta itself.  They may be detected in 
supernova remnants and might account for the evidence of jets in
in Cas A (Fesen \& Gunderson 1996; Reed, Hester \& Winkler 1999).

 The
composition of the jets must  reflect the composition of the innermost
 parts of
the star, and should contain heavy and intermediate-mass elements. 
During the explosion, the jets would bring heavy and intermediate mass elements
into the outer layers. This will influence the spectral and polarization
properties of a supernova.
Here we considered a bare helium core, 
but if the core were inside a hydrogen envelope, 
the explosion would remain very inhomogeneous.
Radioactive elements could potentially be carried into the hydrogen
envelope. This could explain the early appearance of X-rays, 
as in SN1987A.
It is plausible that a sufficiently powerful jet could 
even penetrate a hydrogen envelope.

We assumed that the jets are identical which is not the general case.
Any momentum imbalance might impart a kick to the neutron star. 
From  momentum conservation, we estimate the required difference between
the inflow velocities of the jets, $\Delta v_j$, be of the order of
$$
{\Delta v_j\over v_j} \simeq {M_{NS} \over M_j} \, {v_{NS}\over v_j}
     \simeq 1.0 \, \left( v_{NS}\over 1,000{\rm km/s}\right)\,
                \left( 30,000{\rm km/s}\over v_j \right)~,
$$
where $V_{NS}$ is the kick velocity, and 
we have taken the neutron mass $M_{NS}= 1.5$\m, and the jet mass
 $M_j= 10^{32}~g$. Although the required jet asymmetry,
${\Delta v_j\over v_j}=1$, 
to  produce   a 1,000 km/s kick may seem extreme, the
parameters of jets selected for this calculations are mild. If the duration
of the jets is increased by a factor of two, asymmetry of only 0.5
would be required.

When the jets break through the stellar photosphere, a small amount of 
mass will be accelerated through the density gradient 
to very high velocities.  Our resolution was not
enough and the code does not have a relativistic Riemann solver incorporated
to make quantitative predictions; however, a small fraction of the material
at the stellar surface was observed to move with a   velocity of up to
$\sim$ 90,000 \kms.  This may, in principle, lead to the \grb\ and the 
radio outburst similar to those associated with SN~1998bw/GRB980425.

The jet-induced explosion of a supernova computed in this paper
 is entirely due to
the action of the jet on the surrounding star. 
The mechanism that determines the energy of such an explosion must be
related to the shut-off of the  accretion onto the neutron star
 by the lateral shocks that accelerate the material outwards. 
The explosion thus does not depend 
on neutrino transport or re-acceleration of the stalled shock. 
This work opens many issues that require further investigation. 
A study must be made of 
different input parameters, including properties of the jets and 
of the initial star, and the jet engine mechanisms must be studied as well.
These studies are currently underway.

\acknowledgments
Computations were performed on Origin 2000 at the Naval
 Research Laboratory. 
The authors are grateful to Rob Duncan and
Insu Yi for helpful discussions, and to Almadena Chtchelkanova for 
the development of massively parallel software used in 
the simulations. This research was supported 
in part by NSF Grant 95-28110, NASA Grant NAG 5-2888, NASA Grant LSTA-98-022
and a grant from the Texas Advanced Research Program. 
The Laboratory for Computational Physics and Fluid Dynamics at the Naval
Research Laboratory thanks NASA Astrophysics Theory Program for support.

\newpage

\centerline{\bf Figure Captions}

\figcaption
{Schematic of the simulation. }

\figcaption
{Initial Conditions. The distribution of physical parameters inside the
                      innermost 5\m\ of the 15\m\ stellar model of Straniero
et al. (1999). The Fe-Si inner part is assumed to collapse into a neutron
star.
The O-Ne-Mg, C-O, and He layers are mapped onto the computational domain
(see Section 2).}

\figcaption
{Jet propagation inside the star. The frames show the density in the x-z
plane passing through the center of computational domain. Time since the
beginning
of the simulation is given in the upper left corner of each frame. The sizes
of two upper frames are $\Delta x =7.2\times 10^9$~cm and 
$\Delta z= 9.0\times 10^9$~cm. 
The sizes of the lower frames 
$\Delta x =3.6\times 10^{10}$~cm and 
$\Delta z= 4.5\times 10^{10}$~cm. 
}

\figcaption
{Jet evolution after breakout. The frames show the density in the x-z
plane passing through the center of computational domain. Time since the
beginning
of the simulation is given in the upper left corner of each frame. The sizes
of frames are $\Delta x =6.1\times 10^{10}$~cm and 
$\Delta z= 1.125\times 10^{11}$~cm.
}

\newpage

\plotone{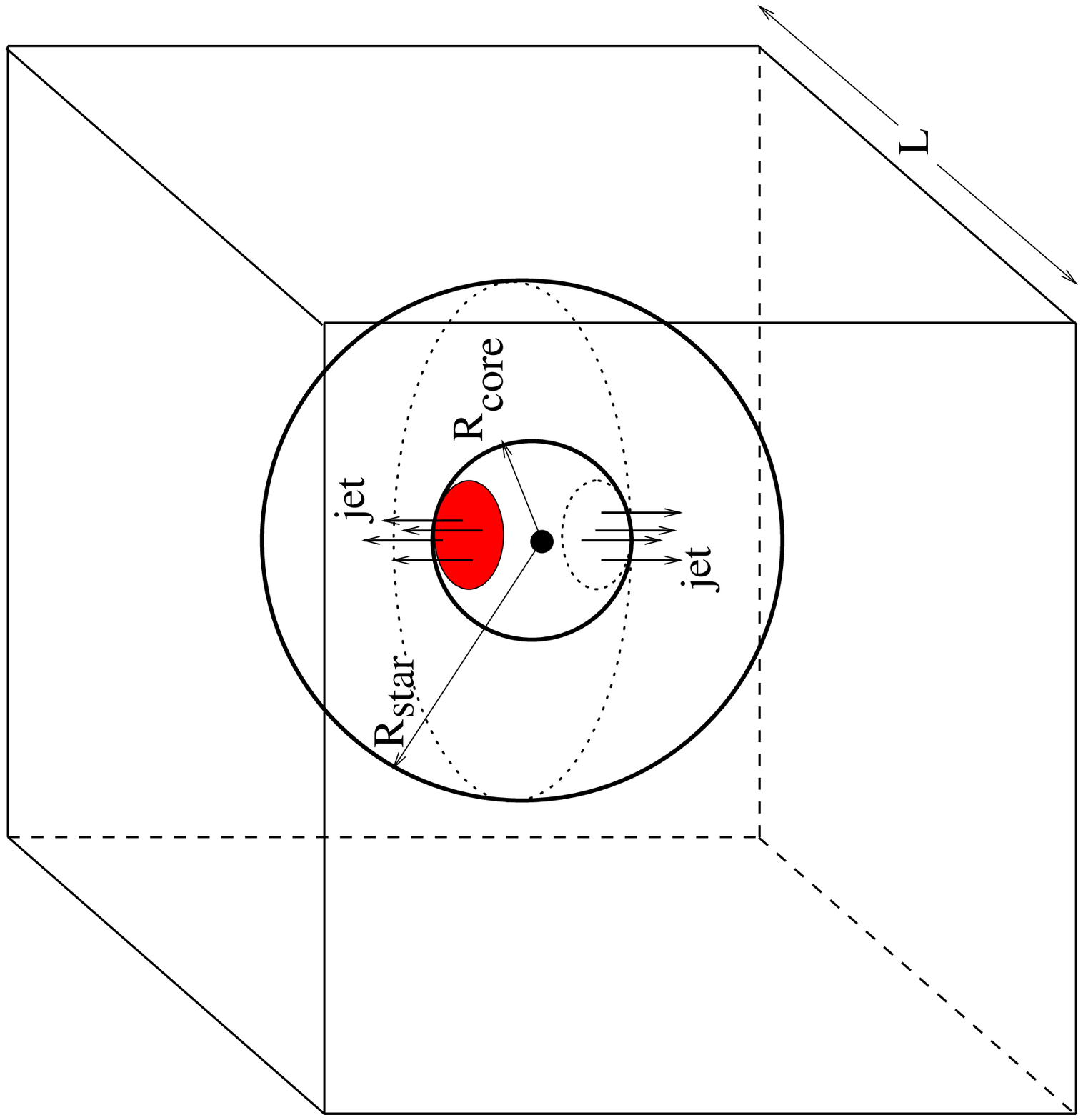}

\bigskip\centerline{\bf Figure 1}

\newpage

\plotone{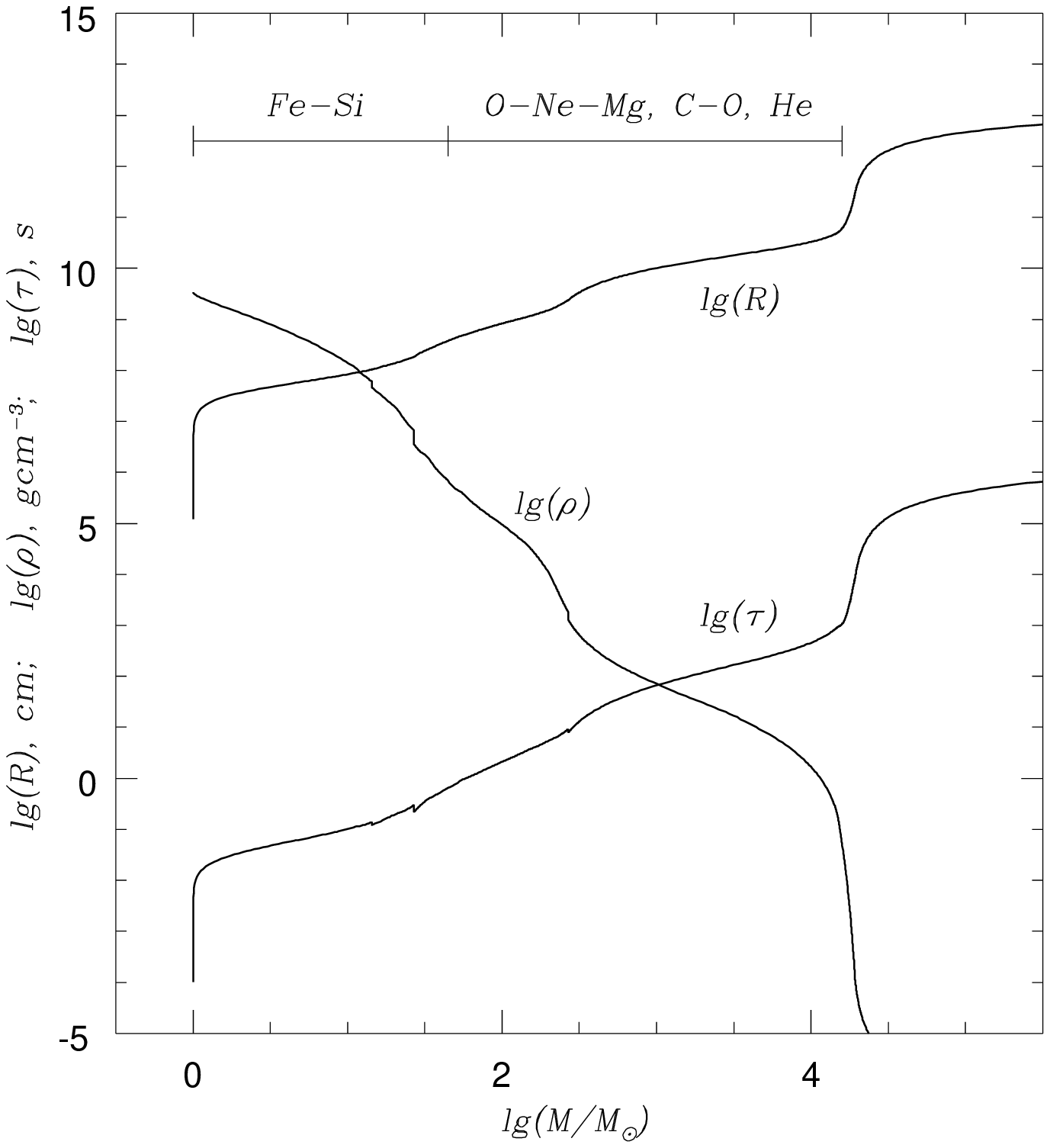}

\bigskip\centerline{\bf Figure 2}

\newpage

\plotone{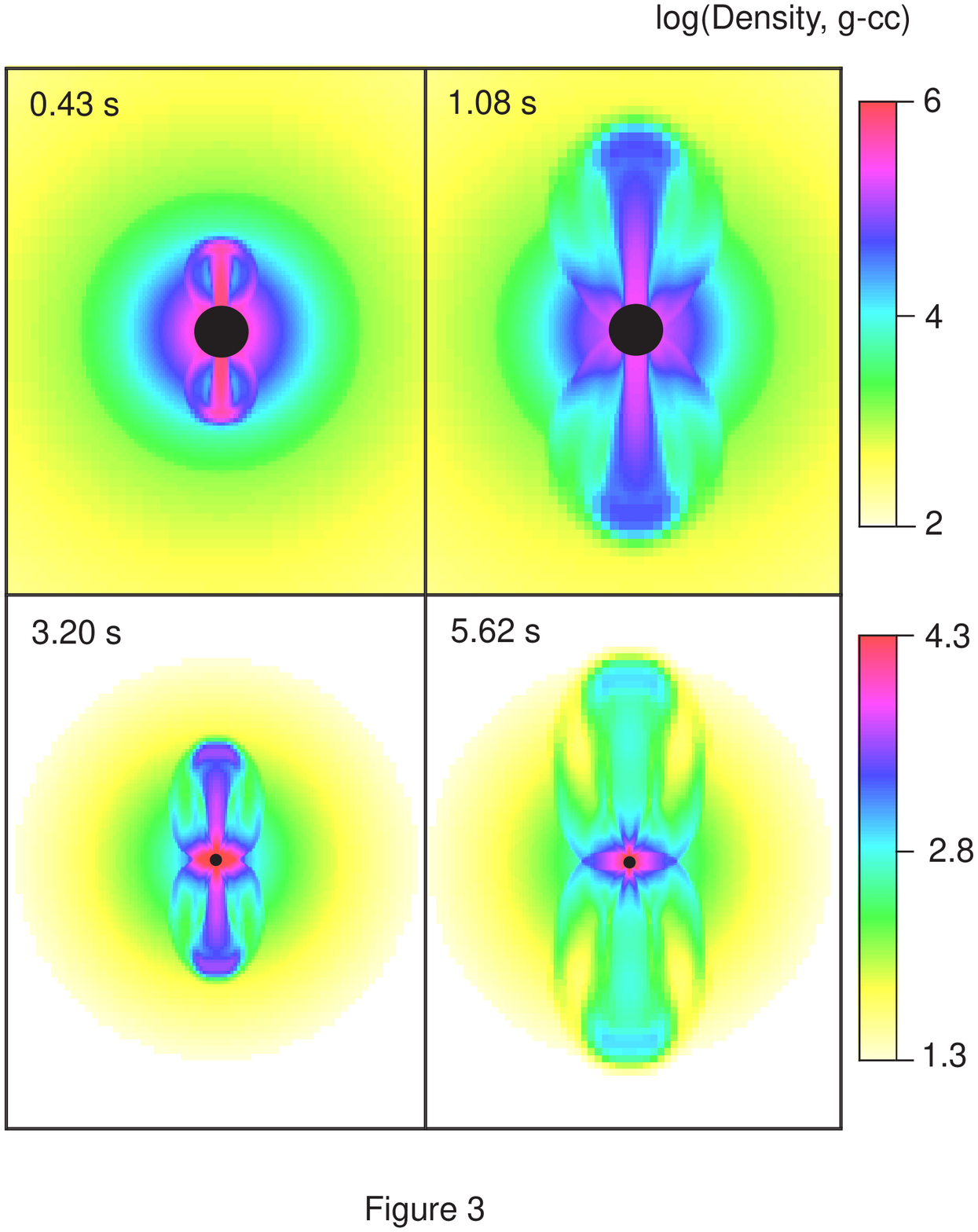}

\newpage

\plotone{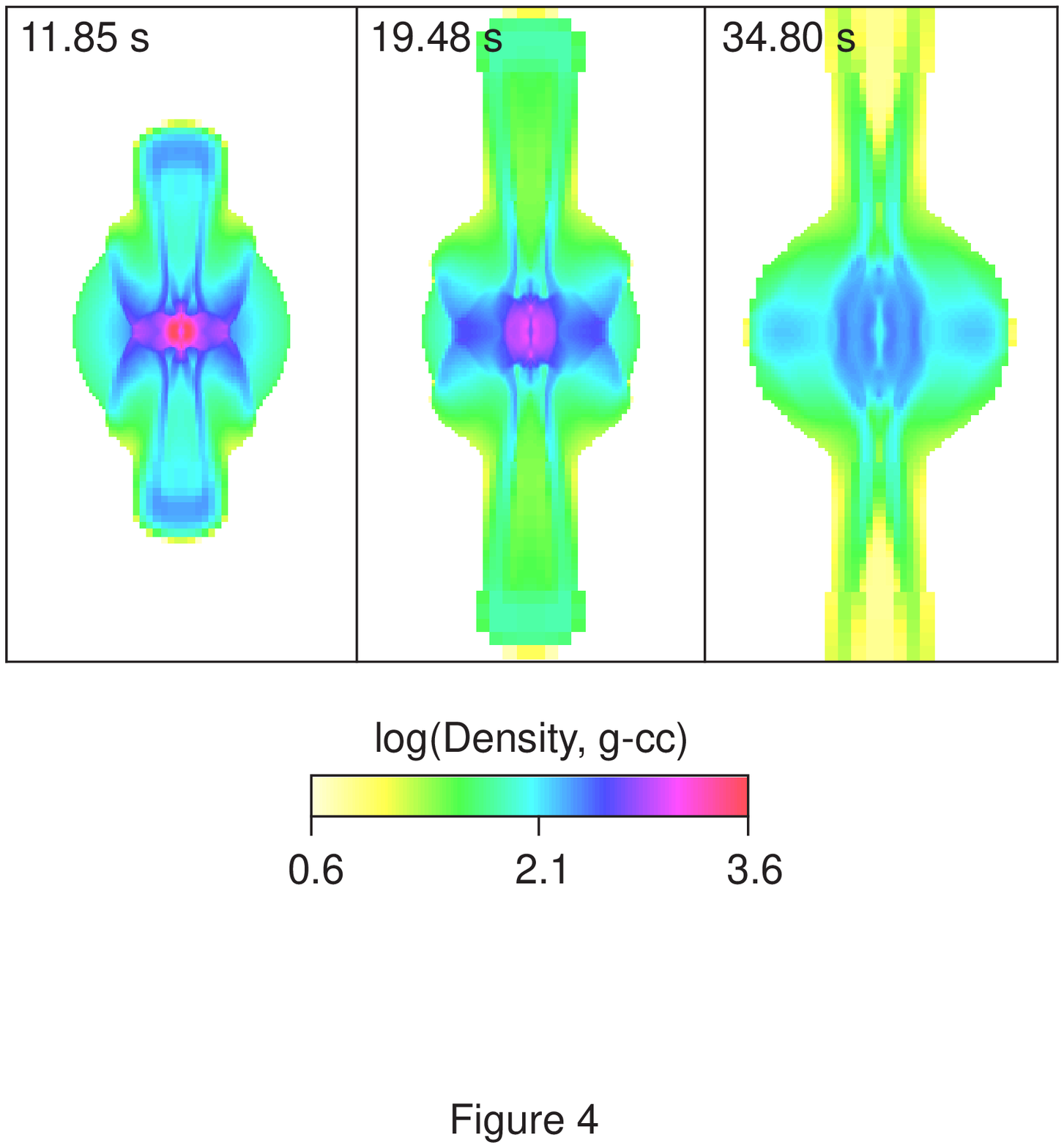}

\end{document}